\documentclass[twocolumn]{revtex4}
\usepackage{amssymb,epsf}
\usepackage{latexsym}
\usepackage{xcolor}
\usepackage{epsfig}
\usepackage{float}
\begin{document}
\title{Hubble tension bounds the GUP and EUP parameters}
\author{S. Aghababaei$^1$\footnote{sarah.aghababaei@gmail.com}, H. Moradpour$^2$\footnote{hn.moradpour@maragheh.ac.ir}, and Elias C. Vagenas$^3$\footnote{elias.vagenas@ku.edu.kw}}
\address{$^1$Department of Physics, Faculty of Sciences, University of Sistan and Baluchestan, Zahedan, Iran\\
$^2$Research Institute for Astronomy and Astrophysics of Maragheh (RIAAM), University of Maragheh, 55136-553, Maragheh, Iran\\
$^{3}$Theoretical Physics Group, Department of Physics,
Kuwait University, P.O. Box 5969, Safat 13060, Kuwait}
\begin{abstract}
\par\noindent
In  recent years the discrepancy in the  value of the Hubble parameter has been growing. Recently, there are works supporting  the  proposal that the uncertainty principles can  solve the Hubble tension. Motivated by this proposal,
we work with an isotropic and homogeneous FRW universe, obtain its Hamiltonian equations, and thus, the Hubble parameter through the first Friedmann equation.  In the context of GUP and EUP models, the Hubble parameter is modified.  Since the fingerprints of quantum gravity are imprinted on the CMB, we consider the GUP/EUP-modified Hubble parameter in the first Friedmann equation to be the one measured by the Planck collaboration which uses the CMB data. The unmodified Hubble parameter in the first Friedmann equation is considered to be the one measured by the HST group which uses the SNeIa data. Therefore, upper bounds for the dimensionless parameters of GUP and EUP are obtained.
\end{abstract}
\maketitle
%
%
%
%
%
%
\section{Introduction}
%
%
%
%
%
\par\noindent
The unification of quantum theory and general relativity is an important issue of contemporary physics. If  quantum mechanics and general relativity are both present in a theory, a critical scale named Planck length $\ell_{p}=\sqrt{\frac{G\hbar}{c^{3}}}\approx10^{-35}~m$, is introduced and considered as the minimum measurable scale of this theory \cite{Maggiore:1993zu}. Furthermore, the existence of such a minimal measurable length is supposed to modify the Heisenberg Uncertainty Principle (HUP) as
\begin{eqnarray}
\Delta X \Delta P\geq \frac{\hbar}{2}\Big[1+\beta (\Delta P)^{2}+\eta (\Delta X)^{2}+\dots\Big]
\label{EGUP}
\end{eqnarray}
\par\noindent 
where $\beta$ and $\eta$ denote the Generalized Uncertainty Principle (GUP) and Extended Uncertainty Principle (EUP) parameters, respectively \cite{Kempf:1994su, Kempf:1996mv, Hassanabadi:2020osz, Bambi:2007ty,Mureika:2018gxl}. There are also other types of GUP 
introduced and investigated, for instance, see Refs.  \cite{Kempf:1993bq,Pedram:2011gw, Pedram:2012my,Das:2008kaa,Das:2010sj,Ali:2009zq}. Different GUPs and EUPs are derived in different contexts such as string theory \cite{Gross:1987ar, Konishi:1989wk}, extra dimensions \cite{Scardigli:2003kr}, black hole physics \cite{Maggiore:1993rv, Scardigli:1999jh}, and propose deep connections with generalized statistics \cite{Moradpour:2019yiq, Shababi:2020evc}.
\par\noindent
On a different note, one of the most significant problems in modern cosmology is the Hubble tension. This is a  disagreement on the current value of Hubble parameter, i.e., $H_0$, between the Planck collaboration which estimates a value  of $H_{CMB}=67.40\pm0.50~\rm km\, s^{-1}Mpc^{-1}$ \cite{Aghanim:2018eyx} and the Hubble Space Telescope (HST) group which gives  $H_{SN}=74.03\pm1.42~\rm km\,s^{-1} Mpc^{-1}$ \cite{Riess:2019cxk}. It is  noteworthy to state that the HST group utilizes SNeIa data while the Planck collaboration employs the Cosmic Microwave Background (CMB) data. This difference has not been eliminated but, on the contrary, during the last years it has been grown  to the  $4.4\sigma$ level \cite{Gonzalez:2020fdy}. 
\par\noindent
Despite the numerous attempts to solve  this discrepancy, its mechanism remains still unclear. There are various proposals and different approaches for solving this problem in cosmology \cite{Vagnozzi:2019ezj,Barker:2020gcp}. In particular, the $H_0$ tension is proposed as the result of factors such as the curvature, the neutrino masses, the effective number of neutrino species, and so on \cite{DiValentino:2017rcr, DiValentino:2017oaw, Mortsell:2018mfj, Yang:2018qmz, Poulin:2018cxd}. As already mentioned, until now there is not a comprehensive agreement on the various proposals of eliminating this discrepancy and therefore, this motivates physicists to think about other possibilities. In this direction, it is considered that during the Planck epoch quantum fluctuations were produced. These quantum fluctuations propagating in spacetime generated the primordial fluctuations in the Inflation era. These primordial fluctuations which are quantified by a power spectrum are encoded in the anisotropies of the CMB \cite{Abazajian:2013vfg, Kaya:2021oih}. So, observing the anisotropies of the CMB, we may ``see" the  fingerprints of Quantum Gravity \cite{Brandenberger:2002hs, Cai:2014hja, Kempf:2018gbt}. Therefore, since  the quantum features of gravity are considered to be stored on CMB \cite{Tsujikawa:2003gh, Akofor:2007fv, Koivisto:2010fk, Krauss:2013pha, Kempf:2018gbt, Calmet:2019tur, Ashtekar:2020gec}, it has been claimed that GUP could solve the Hubble tension problem \cite{Capozziello:2020nyq, uncertitiyHubble2}. 
\par\noindent
In the present work, motivated by the idea that the quantum gravity effects can solve the Hubble tension, we study the consequences of the discrepancy in the value of the Hubble parameter on the GUP and EUP. Specifically, we focus on the modifications of the Hamiltonian of a Friedmann-Robertson-Walker (FRW) universe  due to the different types of GUP and EUP. Therefore, utilizing the different values for the Hubble parameter provided by the Planck collaboration and the HST group, we obtain some bounds on the GUP and EUP parameters. Our results show that  the Hubble tension problem, if it stems from the signals of quantum gravity encoded in the observations, can provide information about the GUP/EUP parameters. In Section II, we  briefly present the Hamiltonian of the FRW universe, derive the Hamiltonian equations in the context of HUP, and thus, find the standard Hubble parameter. In Section III, we generalize the previous analysis in the context of GUP. We obtain the  GUP-modified  Hamiltonian equations, and thus utilizing the $H_0$ tension we derive  a bound for the GUP parameter for two different types of the GUP model. In Section IV,  we work in the context of EUP. We apply a similar  to Section III approach  for the EUP model. We find the corresponding EUP-modified Hamiltonian equations and find a bound for the EUP parameter for two different types of the EUP model. Finally, Section V is devoted to briefly summarize our  results and make some concluding remarks.
%
%
%
%
%
%
\section{cosmological model}
%
%
%
%
\par\noindent
The Hamiltonian of an isotropic and homogeneous FRW universe, in natural units $\hbar=c=16\pi G=1$,  is written in the form \cite{Alvarenga:2001nm, Vakili:2012tm, Ardehali:2016yzg, Poplawski:2016xir,Pedram:2007ud, Rasouli:2014dba, Rasouli:2018nwi}
\begin{eqnarray}
\mathcal{H}_{FRW}(p_{a},a)=N\frac{p_a^{2}}{24a}+6Nka-N\rho a^{3}+\kappa\Pi
\label{FRWhamiltonian}
\end{eqnarray}
\par\noindent 
where $a$ and $p_{a}$ are the scale factor (as the generalized coordinate operator) and the generalized momentum conjugate to the scale factor, respectively.
Additionally, $N$ is the lapse function that has no dynamical role, and $\Pi$ is its conjugate momentum, while $\kappa$ is its corresponding coefficient. It should be noted that for the case of constant $N$, we get $\Pi=0$. Furthermore, $k$ and $\rho$ denote the geometrical parameter of the FRW universe and the density of the universe, respectively.
\par\noindent
Here, we will use the scale factor $a$ instead of the position operator $x$  in the uncertainty relation. Therefore, based on the HUP for the operators $a$ and $p_{a}$, the corresponding Dirac bracket will be  written as  \cite{Alvarenga:2001nm, Vakili:2012tm}
\begin{eqnarray}
[a, p_{a}]=i\hbar~.
\end{eqnarray}
\par\noindent 
At this point, for simplicity reasons we set $N=1$  \cite{Alvarenga:2001nm, Vakili:2012tm,Rasouli:2014dba, Rasouli:2018nwi}. Thus, by employing the Hamiltonian equations, one obtains
\begin{eqnarray}
\dot{a}&=&\frac{\partial \mathcal{H}_{FRW}}{\partial p_{a}}=\frac{1}{12}\frac{p_{a}}{a}~,\nonumber\\
\dot{p}_{a}&=&-\frac{\partial \mathcal{H}_{FRW}}{\partial a}=\frac{1}{24}\frac{p_{a}^{2}}{a^{2}}-6k+3\rho a^{2}+\frac{d\rho}{da}a^{3}~,\nonumber\\
\dot{\Pi}&=&-\frac{1}{24}\frac{p_{a}^{2}}{a}-6ka+\rho a^{3}=0~.
\label{HEq}
\end{eqnarray}
\par\noindent 
Now, we combine the first and third equations of Eq.~(\ref{HEq}), and  so the first Friedmann equation reads
\begin{eqnarray}
H^{2}=\left(\frac{\dot{a}}{a}\right)^{2}= \frac{1}{12^{2}}\frac{p^{2}_{a}}{a^{4}}=\frac{1}{6}\rho-\frac{k}{a^{2}}
\end{eqnarray}
\par\noindent 
which yields
\begin{eqnarray}
H=\sqrt{\frac{1}{6}\rho-\frac{k}{a^{2}}}~.
\label{Hubble}
\end{eqnarray}
\par\noindent 
It is evident that if we consider the case $k=0$ (a flat FRW universe in agreement with the WMAP data \cite{roos}) and 
set $\rho= 2\Lambda$ with $\Lambda$ to be the cosmological constant,  then the Hubble parameter for a cosmological constant-dominated universe will be in the form $H=\sqrt{\frac{\Lambda}{3}}$ \cite{Carroll:2000fy}.
%
%
%
%
%
%
\section{GUP and Hubble parameter}
%
%
%
%
\par\noindent
As already mentioned, fingerprints of quantum gravity are stored in CMB  \cite{Tsujikawa:2003gh, Akofor:2007fv, Koivisto:2010fk, Krauss:2013pha, Kempf:2018gbt, Calmet:2019tur, Ashtekar:2020gec}. Therefore, if CMB data is used to estimate the $H_{0}$ value, as is the case for the Planck collaboration, then quantum gravity corrections have to be taken into consideration. Therefore, following the analysis of the previous section,  Eq.~(\ref{FRWhamiltonian}) should be rewritten in the context of GUP. 
As already mentioned in the Introduction,  there are several versions of GUP which are generalizations of HUP \cite{Gross:1987ar, Konishi:1989wk, Scardigli:2003kr, Maggiore:1993rv, Scardigli:1999jh}. One of the first, and general form, types of GUP is   
\begin{eqnarray}
\Delta X \Delta P\geq \frac{\hbar}{2}\Big(1+ \lambda f(\Delta P)\Big)
\label{GUP}
\end{eqnarray}
\par\noindent 
where $\lambda$ denotes the GUP parameter and $f(P)$ is a function of the particle momentum, i.e., $P$, and/or its  uncertainty, i.e., $\Delta P$. For instance, a well-known type of GUP with only a term quadratic in momentum was given by  Kempf, Mangano, and Mann (KMM type) \cite{Kempf:1994su} 
\begin{eqnarray}
\Delta X \Delta P\geq \frac{\hbar}{2}\Big(1+ \beta \Delta P^{2}\Big)~.
\label{GUPKMM}
\end{eqnarray}
\par\noindent 
An improved version of KMM type of GUP was introduced in Ref.  \cite{Nouicer:2007jg} 
\begin{eqnarray}
\Delta X \Delta P\geq \frac{\hbar}{2} \rm e^{\beta \Delta P^{2}}~.
\label{GUPNouicer}
\end{eqnarray}
\par\noindent
In addition, there is another type of GUP proposed by Pedram \cite{Pedram:2011gw, Pedram:2012my}
\begin{eqnarray}
\Delta X \Delta P\geq \frac{\hbar}{2} \frac{1}{1-\beta P^{2}}~.
\label{GUPPedram}
\end{eqnarray}
\par\noindent 
At this point, a couple of comments are in order. First, at the Planck scale, this type of GUP 
naturally induces a UV cutoff for the momentum. 
Second, this type of GUP has also  been seen in Refs. \cite{Chung:2019raj, Hassanabadi:2019eol, Das:2020ujn}. 
\par\noindent
In all of the above-mentioned types of GUP, the momentum $P_{a}$ (valid near the Planck scale) 
is expressed as a function of the canonical momentum, i.e.,  $p_{a}$, as follows  \cite{Kempf:1994su, Pedram:2011gw, Pedram:2012my, Chung:2019raj, Hassanabadi:2019eol, Das:2020ujn}
\begin{eqnarray}
P_{a}=p_{a}\Big(1+\lambda_{1} p_{a}+\lambda_{2}p_{a}^{2}+\mathcal{O}(p_a^{3}) \Big)
\label{P}
\end{eqnarray}
\par\noindent 
where $\lambda_{1}$ and $\lambda_{2}$ are the GUP parameters of the  corresponding types of GUP  under study. 
\par\noindent
From now on, without loss of generality and  for the sake of simplicity, we will keep terms up to 2nd order in momentum.
Thus, by applying the ``transformation"  $p_a\rightarrow P_a$ in Eq.~(\ref{FRWhamiltonian}), the deformed Hamiltonian of an isotropic and homogeneous FRW universe reads
\begin{eqnarray}
\mathcal{H}_{FRW}^{GUP}(p_{a}, a)&=&\frac{1}{24}\frac{p_{a}^{2}(1+2\lambda_{1} p_{a}+2\lambda_{2}p_{a}^{2}+\lambda_{1}^{2}p_{a}^{2})}{a}+6ka\nonumber\\
&-&\rho a^{3}+\kappa\Pi~.
\label{HGUP}
\end{eqnarray}
\par\noindent 
It is evident that the corresponding GUP-modified Hamiltonian equations will now be of the form
\begin{eqnarray}
\dot{a}=\frac{\partial \mathcal{H}_{FRW}^{GUP}}{\partial p_{a}}
=\frac{1}{12}\frac{p_{a}}{a}(1+3\lambda_{1}p_{a}+4\lambda_{2}p_{a}^{2}+2\lambda_{1}^{2}p_{a}^{2}),
\label{A}
\end{eqnarray}
\begin{eqnarray}
\dot{p}_{a}&=&-\frac{\partial \mathcal{H}_{FRW}^{GUP}}{\partial a}
=\frac{1}{24}\frac{p_{a}^{2}(1+2\lambda_{1}p_{a}+2\lambda_{2}p_{a}^{2}+\lambda_{1}^{2}p_{a}^{2})}{a^{2}}\nonumber\\
&&-\,  6k+3\rho a^{2}+\frac{d\rho}{da}a^{3}~,
\label{pdotGUP}
\end{eqnarray}
\begin{eqnarray}
\dot{\Pi}&=&-\frac{1}{24}\frac{p_{a}^{2}(1+2\lambda_{1}p_{a}+2\lambda_{2}p_{a}^{2}+\lambda_{1}^{2}p_{a}^{2})}{a}-6ka+\rho a^{3}\nonumber\\
&=&0~.
\label{PidotGUP}
\end{eqnarray}
\par\noindent 
By combining  Eqs.~(\ref{A}) and (\ref{PidotGUP}), we obtain
\begin{eqnarray}
H_{GUP}^{2}&=&H^{2}+48\lambda_{1}a^{2}H^{3}+864\lambda_{2}a^{4}H^{4}\nonumber\\
&&+\,576\lambda_{1}^{2}a^{4}H^{4}~.
\label{Hg}
\end{eqnarray}
\par\noindent 
It is obvious that for the case of $\lambda_{1}\neq 0$ and/or $\lambda_{2}\neq 0$, we obtain  $H_{GUP}\neq H$. 
This can be considered as the answer to the Hubble tension problem. To be more specific, on one hand, as already mentioned,  CMB carries fingerprints of quantum gravity \cite{Tsujikawa:2003gh, Akofor:2007fv, Koivisto:2010fk, Krauss:2013pha, Kempf:2018gbt, Calmet:2019tur, Ashtekar:2020gec}. Therefore, the Hubble parameter reported by the Planck collaboration  ($H_{CMB}$) can be viewed as the GUP-modified Hubble parameter  in Eq. (\ref{Hg}), i.e., $H_{CMB}=H_{GUP}$. On the other hand, the supernova phenomena are explained by quantum mechanics quite well, and furthermore,  based on our current understanding, general relativity is adequate to study the spacetime around supernovae. Therefore, we can view the Hubble parameter  reported by the HST group as the unmodified Hubble parameter in Eq. (\ref{Hg}), i.e.,  $H_{SN}=H$.
\par\noindent
First, we consider the case $\lambda_{2}=0$ which can be found in Refs.~\cite{Chung:2019raj, Hassanabadi:2019eol, Das:2020ujn}. From Eq. (\ref{Hg}), the  Hubble parameter now reads
\begin{eqnarray}
H_{CMB}=H_{SN}\Big(1+24\lambda_{1}a^{2}H_{SN}\Big)~.
\label{HGUP1}
\end{eqnarray}
\par\noindent 
At this point, a couple of comments are in order.
First, the above expression implies that $\lambda_{1}$ has to be negative since $H_{CMB}< H_{SN}$ according to the recent observations.  This is in agreement  with Refs. \cite{Das:2008kaa,Das:2010sj,Ali:2009zq} where  the linear term in momentum is subtracted from unity. 
Second, by using Eq.~(\ref{HGUP1}), one can  easily obtain an expression for $\lambda_{1}$ as
\begin{eqnarray}
\frac{H_{CMB}-H_{SN}}{H_{SN}^{2}}=24\lambda_{1}a^{2}~.
\label{a1}
\end{eqnarray}
\par\noindent
By  considering the recent reports for the values of the Hubble parameter, i.e., $H_{CMB}$ and $H_{SN}$, and setting  $a=1$ for today, we find  $|\lambda_{1}|\approx 1.5 \times 10^{40}~\rm GeV^{-1}$. 
 At this point, we need to reinstate the units in Eq. (\ref{a1}) and  we write  $\lambda_{1}=\frac{\lambda_{01}}{M_p c}$, with  $\lambda_{01}$ to be the dimensionless GUP parameter and  $M_p$ to be the Planck mass. Therefore, we obtain the value for the dimensionless GUP parameter  $|\lambda_{01}| \approx 2.9 \times 10^{58}$ which will be an upper bound for Eq.~(\ref{a1}) since future observations will reduce the discrepancy. 
 \par\noindent
Second, we consider the case $\lambda_{1}=0$ which can be found in Ref. \cite{Kempf:1993bq}. From Eq. (\ref{Hg}), the  Hubble parameter now reads
\begin{eqnarray}
H_{CMB}\simeq H_{SN}\Big(1+432\lambda_{2}a^{4}H_{SN}^{2}\Big)~.
\label{19}
\end{eqnarray}
\par\noindent
At this point, a couple of comments are in order.
First, the above expression implies that $\lambda_{2}$ has to be negative since $H_{CMB}< H_{SN}$ according to the recent observations. This is in agreement  with the negative dimensionless GUP parameter obtained in Ref. \cite{Scardigli:2014qka}. Second, by using Eq.~(\ref{19}), one can  easily obtain an expression for $\lambda_{2}$ as
\begin{eqnarray}
\frac{H_{CMB}-H_{SN}}{H_{SN}^{3}}=432\lambda_{2}a^{4}~.
\label{a2}
\end{eqnarray}
\par\noindent 
By considering the recent reports for the values of the Hubble parameter, i.e., $H_{CMB}$ and $H_{SN}$, and setting  $a=1$ for today, we find  $|\lambda_{2}|\approx 3.7 \times 10^{42}~\rm GeV^{-2}$. 
By reinstating the units in Eq. (\ref{a1}), we write  the GUP parameter as $\lambda_{2}=\frac{\lambda_{02}}{(M_p c)^{2}}$, with  $\lambda_{02}$ to be the dimensionless GUP parameter and  $M_p$ to be the Planck mass. Therefore, we obtain the value for the dimensionless GUP parameter  $|\lambda_{02}| \approx 1.3 \times 10^{79}$ which will be an upper bound for Eq.~(\ref{19}), namely for the dimensionless GUP parameters, since that future observations will reduce the discrepancy. 
%
%
%
%
%
%
%
%
\section{EUP and Hubble parameter}
%
%
%
%
%
%
%
%
\par\noindent
In this section, we will employ a similar to the previous section analysis but for the case of the EUP model. It is known that for cosmological models the scale factor $a$ plays the role of the position  $x$. In the context of the EUP model, the momentum operator (valid near the Planck scale) reads
\begin{eqnarray}
P_a=p_a(1+\eta_{1}a+\eta_{2}a^{2})
\label{PEUP}
\end{eqnarray}
\par\noindent 
where $\eta_{1}$ and $\eta_{2}$ are the dimensionless EUP parameters. 
From now on, without loss of generality and  for the sake of simplicity, we will keep terms up to 2nd order in position.
Thus, by applying the ``transformation"  $p_a\rightarrow P_a$ in Eq.~(\ref{FRWhamiltonian}), the deformed Hamiltonian of an isotropic and homogeneous FRW universe reads
\begin{eqnarray}
\mathcal{H}_{FRW}^{EUP}(p_a, a)&=&\frac{1}{24}\frac{p_a^{2}(1+2\eta_{1}a+\eta_1^{2} a^{2}+2\eta_{2} a^{2})}{a}+ 6ka\nonumber \\
&&  -\, \rho a^{3}+\lambda\Pi ~.
\label{HEUP}
\end{eqnarray}
\par\noindent 
It is evident that the corresponding GUP-modified Hamiltonian equations will now be of the form
\begin{eqnarray}
\dot{a}=\frac{\partial \mathcal{H}_{FRW}^{EUP}}{\partial p_a}=\frac{1}{12}\frac{p_{a}}{a}(1+2\eta_{1}a+\eta_1^{2} a^{2}+2\eta_{2} a^{2})~,
\label{aEUP}
\end{eqnarray}
\begin{eqnarray}
\dot{p}_a&=&-\frac{\partial \mathcal{H}_{FRW}^{EUP}}{\partial a}=\frac{1}{24}\frac{p_a^{2}}{a^{2}}(1-\eta_1^{2} a^{2}-2\eta_{2}a^{2})-6k\nonumber\\
&& +\, 3\rho a^{2}+\frac{d\rho}{da}a^{3}~,
\label{bEUP}
\end{eqnarray}
\begin{eqnarray}
\dot{\Pi}&=&-\frac{1}{24}\frac{p_a^{2}}{a}(1+2\eta_{1}a+\eta_1^{2} a^{2}+2\eta_{2}a^{2})-6ka+\rho a^{3}\nonumber\\
&=&0~.
\label{cEUP}
\end{eqnarray}
\par\noindent
By combining Eqs. (\ref{aEUP}) and (\ref{cEUP}), we obtain
\begin{eqnarray}
H_{EUP}^{2}=H^{2}+2\eta_{1}aH^{2}+\eta_{1}^{2}a^{2}H^{2}+2\eta_{2}a^{2}H^{2}~.
\label{EUPH2}
\end{eqnarray}
Now, if we consider the cases  $\eta_{1}=0$ and $\eta_{2}=0$, we get, respectively,  
\begin{eqnarray}
H_{EUP}= H(1+\eta_{1}a)~,
\label{HEUP2}
\end{eqnarray}
\begin{eqnarray}
H_{EUP}\simeq H(1+\eta_{2}a^{2})~.
\label{HEUP22}
\end{eqnarray}
\noindent 
Utilizing Eq. (\ref{HEUP2}) and Eq. (\ref{HEUP22}), we can express the dimensionless GUP parameters, i.e., $\eta_{1}$ and $\eta_{2}$, in terms of the Hubble parameter as follows, respectively,
\begin{eqnarray}
\frac{H_{CMB}-H_{SN}}{H_{SN}}=\eta_{1}a ~,
\label{a3}
\end{eqnarray}
\begin{eqnarray}
\frac{H_{CMB}-H_{SN}}{H_{SN}}=\eta_{2}a^{2}~.
\label{a4}
\end{eqnarray}
\par\noindent
 Now, by considering recent reports for the values of the Hubble parameter, i.e., $H_{CMB}$ and $H_{SN}$, and setting  $a=1$ for today, dimensionless GUP parameters are both negative  and  their magnitudes are  $|\eta_{1}|=|\eta_{2}|=  9.0\times10^{-2}$. This value  can be considered as  an upper bound for Eq.~(\ref{a3}) and  Eq.~(\ref{a4}), namely for the dimensionless EUP parameters, due to the fact that future observations will reduce the discrepancy. 
%
%
%
 %
 %
 %
%
			
%
%
%
%
%
%
\section{Conclusion}
%
%
%
%
%
\par\noindent
In  recent years the Hubble tension has been growing. Recently, many  works supporting  the  proposal that the uncertainty principles can  solve the Hubble tension \cite{Capozziello:2020nyq, uncertitiyHubble2}. Following this proposal, we exploit the discrepancy between two different ways of measuring the rate of the universe's expansion, to derive bounds on the dimensionless  GUP/EUP parameters. Specifically, in this work, we consider an isotropic and homogeneous FRW universe, and relying on canonical variables, i.e.,  $a$ and $p_a$, we obtain the first Friedmann equation. Then, it is shown that when GUP/EUP is taken into account, the Hubble parameter is GUP/EUP-modified. 
By adopting the proposal that  quantum gravity effects are stored in the CMB, we consider the GUP/EUP-modified Hubble parameter to be the one measured by the Planck collaboration which used the CMB data, while the unmodified Hubble parameter is considered to be the one measured by the HST group which used the SNeIa data. 
Finally,  since we expect future observations to reduce this discrepancy,  the values for the dimensionless GUP/EUP parameters obtained with recent data can be   the upper bounds for these GUP/EUP parameters. In particular, the dimensionless GUP parameter $\lambda_{01}$ is similar to the dimensionless GUP parameter $\alpha_{0}$ introduced in Refs. \cite{Das:2008kaa,Das:2010sj,Ali:2009zq}. Thus, the bound obtained here, namely $|\lambda_{01}| < 2.9\times 10^{58}$ is a not-so-tight bound compared to the ones existing  in the literature, for instance, see Refs. \cite{Das:2009hs,Das:2021lrb}.
The dimensionless GUP parameter $\lambda_{02}$ is similar to the dimensionless GUP parameter $\beta_0$ \cite{Kempf:1994su}. Thus, the bound obtained here, namely $|\lambda_{02}| < 1.3\times 10^{79}$, is similar to the bounds achieved using data from our Solar system, for instance, see Ref. \cite{Scardigli:2014qka}. 
The dimensionless EUP parameters, namely $\eta_{1}$ and $\eta_{2}$, introduce a bound $|\eta_{1}|=|\eta_{2} |<  9.0\times 10^{-2}$ which is a very tight bound compared to the ones existing in the literature,  for instance, see Ref. \cite{Roushan:2019miz}.\\
\vspace{4cm}\\
{\bf Data Availability Statement:}\\
 %
Data sharing not applicable to this article as no datasets were generated or analyzed during the current study.

 %
 %
 %
 %
 %
 %
%
%

\end{document}